\documentclass[epj,twocolumn]{webofc}
\usepackage{graphicx}
\usepackage[varg]{txfonts}   
\usepackage[english]{babel}
\usepackage{multicol}
\usepackage{paralist}
\usepackage{setspace}
\usepackage{sidecap}
\usepackage{palatino}
\usepackage{fancyvrb}
\usepackage{color}
\usepackage[T1]{fontenc}
\usepackage[utf8]{inputenc}
\usepackage{amsmath}
\usepackage{amssymb}
\usepackage{amsfonts}
\usepackage{setspace}
\usepackage{listings}
\usepackage{slashed}
\usepackage[nice]{units}
\usepackage{rotating}
\usepackage{natbib}

\newcommand{\xmm}{\textit{XMM-Newton}}
\newcommand{\fermi}{\textit{Fermi/LAT}}
\newcommand{\swift}{\textit{Swift}}
\newcommand{\pks}{\mbox{PKS\,2004$-$447}}
\newcommand{\kev}{\mathrm{keV}}
\newcommand{\sun}{\odot}

\begin{document}
\woctitle{The Innermost Regions of Relativistic Jets and Their Magnetic Fields}

\title{\center X-ray monitoring of the radio and $\gamma$-ray loud Narrow-Line Seyfert 1 Galaxy PKS\,2004$-$447}
\author{A. Kreikenbohm\inst{1,2}\fnsep\thanks{\email{akreikenbohm@astro.uni-wuerzburg.de}} \and
        M. Kadler\inst{1}
              \and
        J. Wilms\inst{2}
              \and
	R. Schulz\inst{1,2}
             \and
        C. M\"{u}ller\inst{2,1}
              \and
	R. Ojha\inst{3}
             \and
        E. Ros\inst{4,5,6}
	     \and
        K. Mannheim\inst{1}
	     \and
	D. Els\"{a}sser\inst{1}
}

\institute{Lehrstuhl f\"{u}r Astronomie, Universit\"{a}t W\"{u}rzburg, W\"{u}rzburg, Germany 
\and
           Dr. Remeis Sternwarte \& ECAP, Universit\"{a}t Erlangen-N\"{u}rnberg, Bamberg, Germany
\and	
	   ORAU/NASA/GSFC Greenbelt, MD, USA
\and		
	   Observatori Astron\`{o}mic, Univ. Val\`{e}ncia, Spain 
	   \and
	   Dept. Astronom\`{i}a y Astrof\`{i}sica, Univ. Val\`{e}ncia, Spain
\and
           Max-Planck-Institut f\"{u}r Radioastronomie, Bonn, Germany
}

\abstract{
 We present preliminary results of the X-ray analysis of \xmm{} and \swift{} observations as part of a multi-wavelength monitoring campaign in 2012 of the radio-loud narrow line Seyfert 1 galaxy \pks{}. The source was recently detected in $\gamma$-rays by \fermi{} among only four other galaxies of that type. The 0.5$-$10\,$\mathrm{keV}$ X-ray spectrum is well-described by a simple absorbed powerlaw ($\Gamma\,\sim 1.6$). The source brightness exhibits variability on timescales of months to years with indications for spectral variability, which follows a ``bluer-when-brighter`` behaviour, similar to blazars.
}

\maketitle
\section{Introduction}
\label{sec-int}
The recent detection of variable $\gamma$-ray emission in five radio-loud Narrow-Line Seyfert 1 (RL-NLS1) galaxies \cite{Abdo2009a} suggests that these sources possess powerful relativistic jets, similar to blazars and ''classical'' radio galaxies. Narrow line Seyfert 1 (NLS1) galaxies are defined by the presence of narrow permitted optical emission lines, i.e., a strong but narrow H$\beta$ emission and a strong Fe\,\textsc{II} emission \cite{Osterbrogge}. Typically Seyfert 1 galaxies are variable on timescales of months to years. On the contrary, NLS1 exhibit very strong X-ray variability in flux and spectral shape on shorter timescales down to weeks (e.g., \cite{BollerBrandtFabian1997MNRAS289}) and a strong soft excess below $2\,\kev$. These timescales are common for radio-loud active galactic nuclei (AGN) and are associated with emission of the jet and are accompanied by flaring in other wavelengths, from radio to $\gamma$-rays. The detection of $\gamma$-ray emitting RL-NLS1 present a unique opportunity to study the similarities and differences in the emission processes and structure of radio-loud and quiet AGN, since these sources show properties of both types.

The active galaxy \pks{} is a $\gamma$-ray loud NLS1 galaxy at a redshift of $z=0.24$ \cite{Drinkwater}. First observations yield optical
 properties consistent with the formal NLS1 classification, \mbox{(flux ratio O\,[\textsc{III}]/H$\beta=1.6$, H$\beta_\mathrm{FWHM}=1447\,\mathrm{km\,s}^{-1}$)}, but exhibit atypical weak Fe\,\textsc{II} emission \mbox{($EW_{Fe\mathrm{\,\textsc{II}}}\le10\AA$)}
 \cite{Oshlack2001}. The black hole mass is estimated to be approximately $10^{7}\,M_\sun$ \cite{Oshlack2001}. \pks{} was detected in $\gamma$-rays by the \fermi{} Telescope \cite{Abdo2009a} in 2009. Following its discovery in $\gamma$-rays it has been included in the  Very Long Baseline Interferometry (VLBI) TANAMI program \cite{Ohja2010} and is monitored since October 2010. \\
As of the time of writing, among the five known $\gamma$-NLS1 detected in 2009, only one multi-wavelength campaign has been performed on \mbox{PMN\,J0948$+$0022} \cite{Abdo2009a}. However, \pks{} differs dramatically from \mbox{PMN\,J0948$+$0022}, in terms of both physical and spectral properties, such as luminosity and black hole mass \cite{Abdo2009a}. 

\begin{table*}[t]
\caption{Results of the absorbed powerlaw fit to the X-ray data.}
\label{tab-bfpar-abspow}
\centering\small
\begin{tabular}{lccccccr}
\hline \hline
ObsID$^\mathrm{a}$  & ObsDate$^\mathrm{b}$  & Inst $^\mathrm{c}$  &     $\Gamma^\mathrm{d}$             & $F_{0.5-10\,\kev}^\mathrm{e,1}$         & $F_{0.5-2\,\kev}^\mathrm{e,2}$                 &  $F_{2-10\,\kev}^\mathrm{e,3}$     & Stat/dof (red)$^\mathrm{f}$ \\    
\hline
0003249200[5,6] &2013-07-07& XRT& $1.49_{-0.13}^{+0.10}$ & $1.07_{-0.12}^{+0.14}$ & $0.29_{-0.03}^{+0.03}$ & $0.79_{-0.12}^{+0.14}$ & $86.3$/$77$ (1.12)	\\
0694530201  	&2012-10-18& EPIC&$1.67_{-0.05}^{+0.04}$ & $0.61_{-0.03}^{+0.03}$ & $0.20_{-0.01}^{+0.01}$ & $0.41_{-0.03}^{+0.03}$ & $111.9$/$102$ (1.10)	\\
00032492004 	&2012-09-30& XRT& $1.13_{-0.33}^{+0.43}$ & $0.83_{-0.33}^{+0.07}$ & $0.15_{-0.04}^{+0.05}$ & $0.69_{-0.34}^{+0.66}$ & $8.3$/$9$ (0.92)	\\
00032492003 	&2012-09-01& XRT& $2.02_{-0.43}^{+0.44}$ & $0.44_{-0.11}^{+0.16}$ & $0.20_{-0.05}^{+0.06}$ & $0.24_{-0.11}^{+0.17}$ & $12.9$/$9$ (1.43)	\\
00032492001 	&2012-07-03& XRT& $1.79_{-0.41}^{+0.41}$ & $0.56_{-0.16}^{+0.24}$ & $0.21_{-0.05}^{+0.06}$ & $0.36_{-0.15}^{+0.24}$ & $7.1$/$10$ (0.71)	\\
0694530101  	&2012-05-01& EPIC&$1.68_{-0.05}^{+0.06}$ & $0.43_{-0.03}^{+0.03}$ & $0.14_{-0.01}^{+0.01}$ & $0.29_{-0.02}^{+0.03}$ & $111.3$/$87$ (1.27)	\\
00091031007 	&2012-03-14& XRT& $1.43_{-0.32}^{+0.36}$ & $0.45_{-0.15}^{+0.25}$ & $0.11_{-0.03}^{+0.04}$ & $0.34_{-0.15}^{+0.26}$ & $8.5$/$9$ (0.94)	\\
00091031006 	&2011-11-15& XRT& $1.90_{-0.45}^{+0.40}$ & $0.40_{-0.11}^{+0.19}$ & $0.16_{-0.04}^{+0.04}$ & $0.24_{-0.11}^{+0.19}$ & $5.8$/$11$ (0.52	\\
00091031005 	&2011-09-17& XRT& $1.49_{-0.21}^{+0.23}$ & $1.12_{-0.21}^{+0.25}$ & $0.30_{-0.05}^{+0.06}$ & $0.83_{-0.21}^{+0.26}$ & $43.3$/$27$ (1.6)	\\
00091031002 	&2011-07-14& XRT& $1.51_{-0.46}^{+0.45}$ & $0.64_{-0.21}^{+0.35}$ & $0.18_{-0.05}^{+0.06}$ & $0.47_{-0.21}^{+0.36}$ & $16.0$/$8$ (2.0)	\\
00091031001 	&2011-05-15& XRT& $1.67_{-0.41}^{+0.41}$ & $0.40_{-0.12}^{+0.18}$ & $0.13_{-0.04}^{+0.05}$ & $0.28_{-0.12}^{+0.18}$ & $6.2$/$9$ (0.68)	\\
0200360201  	&2004-04-11& EPIC& $1.53_{-0.03}^{+0.02}$ & $1.31_{-0.04}^{+0.04}$ & $0.37_{-0.01}^{+0.01}$ & $0.94_{-0.04}^{+0.04}$ & $222.1$/$258$(0.86)	\\
\hline
\end{tabular}
\begin{minipage}{\textwidth}\small
\textbf{Notes}: Best-fit parameters of the absorbed powerlaw. The absorption is fixed to its galactic absorption, $3.17\times 10^{20}\,\mathrm{cm}^{-2}$ \cite{Kalberla}. Uncertainties correspond to the 90\% confidence limits.
$^\mathrm{a}$ Observation ID, [n,m] marks merged spectra.
$^\mathrm{b}$ Date of observation.
$^\mathrm{c}$ X-ray Instrument. EPIC includes both the pn and MOS 1 \& 2 cameras.
$^\mathrm{d}$ Photon Index of the powerlaw.
$^\mathrm{e}$ Unabsorbed fluxes in units of $10^{-12}\,\mathrm{erg\,cm}^{-2}\,\mathrm{s}^{-1}$ for the $^\mathrm{1}$complete, $^\mathrm{2}$soft  and $^\mathrm{3}$hard energy ranges, respectively.
$^\mathrm{f}$ $\chi^2$ [C-value] per degrees of freedom (dof) and reduced value for \xmm{} [\swift{}] data.
\end{minipage}
\end{table*}

\section{Observations}
\label{sec-obs}
From May to October 2012, we conducted a multiwavelength monitoring program of \pks{} including \xmm{} and \swift{} observations, providing optical/UV and X-ray data coverage, as well as quasi-simultaneous VLBI observations by TANAMI. The X-ray monitoring consists of two deep \xmm{} pointings of $\sim$40\,ks on 2012-05-01 and 2012-10-18, which were connected by three \swift{} observations of approximately 7\,ks. A follow-up observation was also performed 2013-07-07 by \swift{}. In order to to study the long-term behaviour of \pks{} archival data of \xmm{} and \swift{} have been included in the analysis. Here, we present first results of the data analysis of \xmm{} EPIC pn/MOS and \swift{} XRT (see Table \ref{tab-bfpar-abspow}). 

\section{X-ray Analysis}
\label{sec-ana}
Each spectrum was fitted individually with a absorbed powerlaw in the $0.5-10\,\kev$ energy range.For the galactic absorption a value of $N_\mathrm{H}=3.17\times 10^{20}\,\mathrm{cm}^{-2}$ was assumed, based on the LAB Survey \cite{Kalberla}. 
\xmm{} EPIC pn/MOS spectra of each observation were rebinned to a signal-to-noise ratio (SNR) of 5, to ensure the use of $\chi^2$-statistics. \swift{} XRT spectra were rebinned to at least 5 spectral data counts per energy bin and evaluated with Cash statistics \cite{Cash1975}, due to their small SNR. The model yields a good fit (see Table \ref{tab-bfpar-abspow}). Adding freely variable excess absorption did not improve the fit significantly. Thus, to reduce the number of free parameters $N_\mathrm{H}$ was fixed to its galactic value. We tested for the existence of an  Fe\,K$\alpha$ emission line by adding a narrow line at 6.4\,keV to the \xmm{} data. However, it is statistically not significant ($\Delta\chi^2=0.1$ for $\Delta$dof$=1$) and yields an equivalent width of $\mathrm{EW}_{6.4\kev}\le60\,\mathrm{eV}$ in all three observations. A soft excess cannot be confirmed in our data.

\section{Results}
\label{sec-res}
\begin{figure*}[t]
  \centering
  \includegraphics[width=\textwidth]{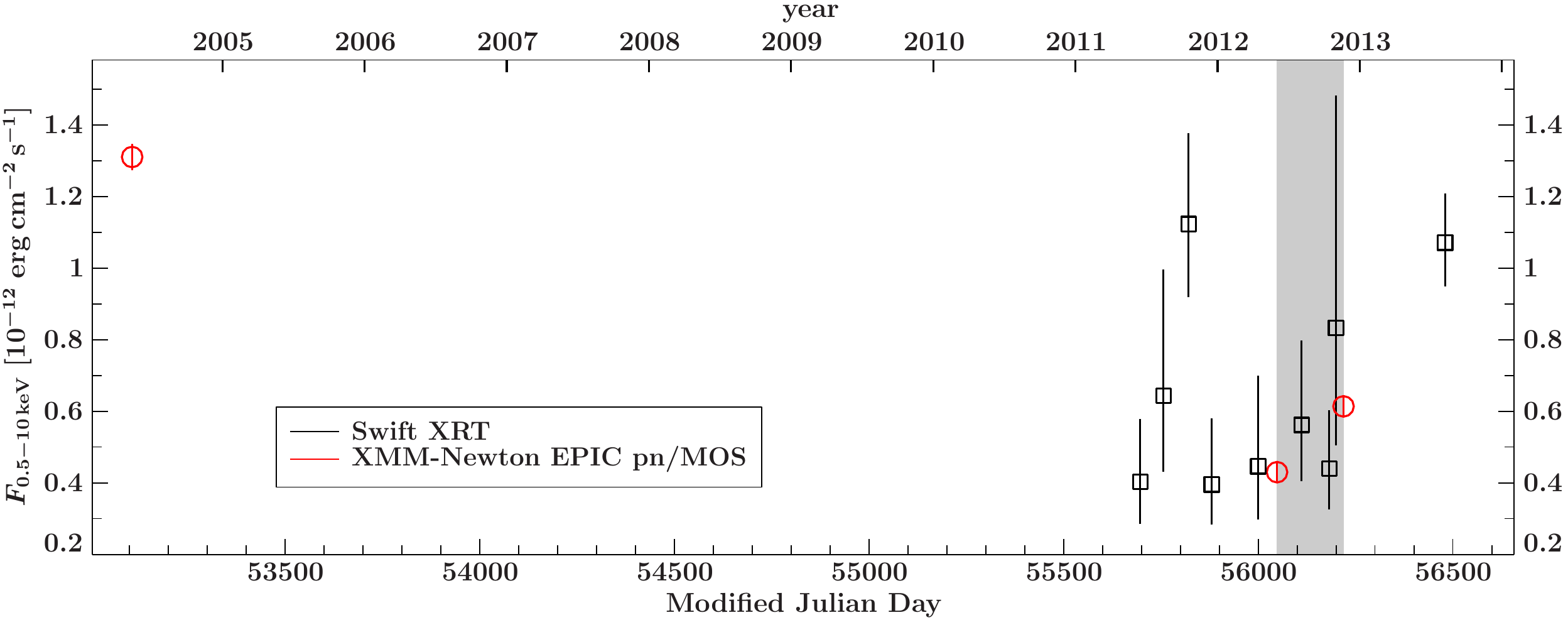} 
  \caption{Unabsorbed $0.5-10\,\kev$ flux of \pks{} as a function of time. Red dots refer to \xmm{} observations, while black suqares resemble \swift{} XRT observations. The grey-shaded background marks the period of our multiwavelength campaign.}
  \label{flux}
\end{figure*}

Figure \ref{flux} shows results for the unabsorbed fluxes of all X-ray observations. A significant decrease of 50\% in flux is observed between 2004 April, and the recent observations in 2011 and 2012. Furthermore, among observations in $2011-2013$ there are indications for flux variations on much smaller timescales. During the monitoring (grey-shaded period), the flux increased by 50\% from 2012-05-01 to 2012-10-18 and, in the following, up to $(1.07\pm0.1)\times10^{-12}\,\mathrm{erg\,cm}^{-2}\mathrm{s}^{-1}$ until July, 2013, which corresponds to a total variation of approximately $0.64\times10^{-12}\,\mathrm{erg\,cm}^{-2}\mathrm{s}^{-1}$ in 14 months. This high flux state is similar to the one in 2004. Another possible rise in flux may be seen in the swift observation from 2011-09-17, where the flux increased and decreased again by approximately 50\% in two months, respectively. 

\begin{figure*}[t]
\centering
  \begin{minipage}{0.5\textwidth}
    \includegraphics[width=\linewidth]{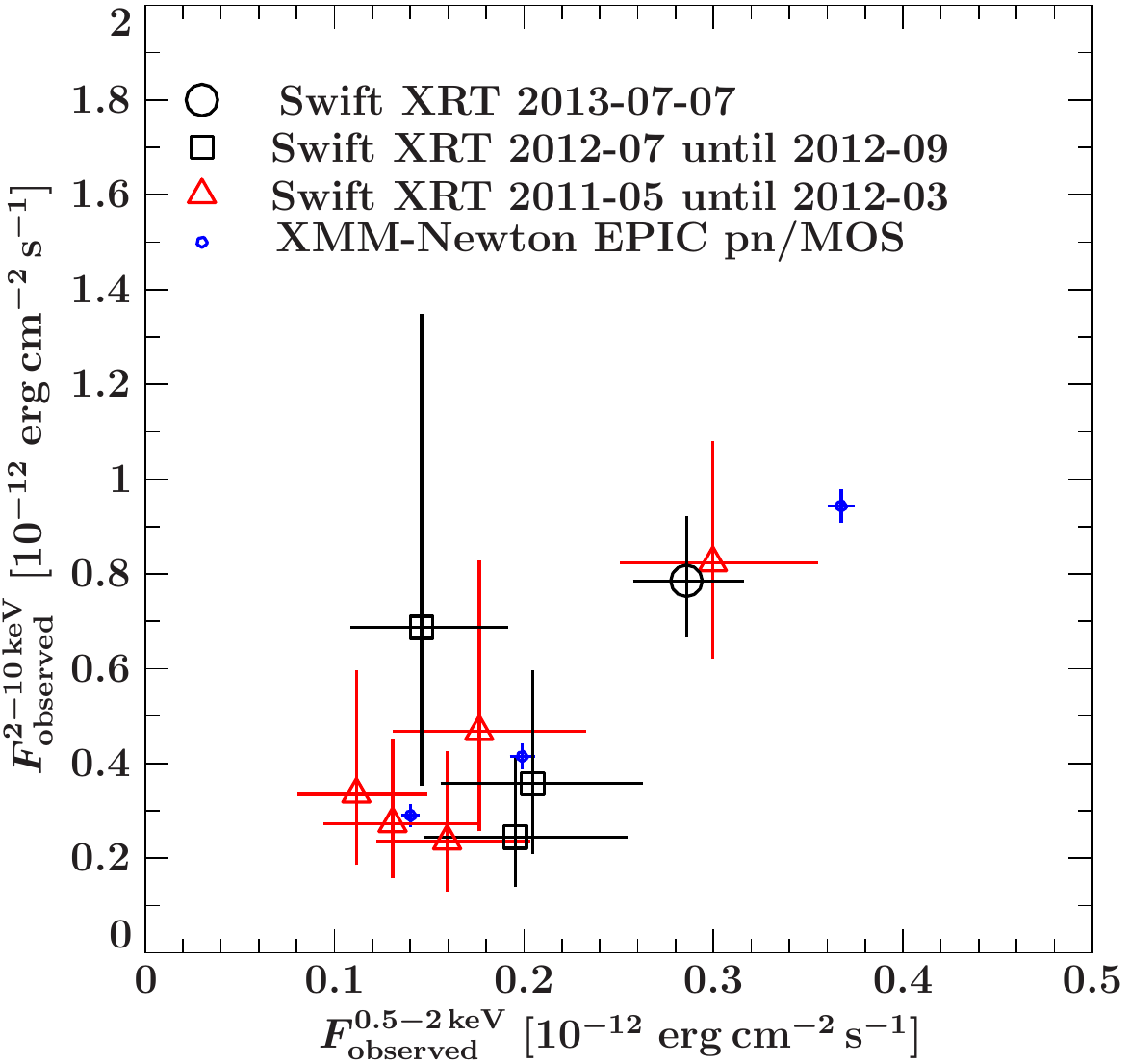} 
  \end{minipage}\begin{minipage}{0.5\textwidth}
    \includegraphics[width=\linewidth]{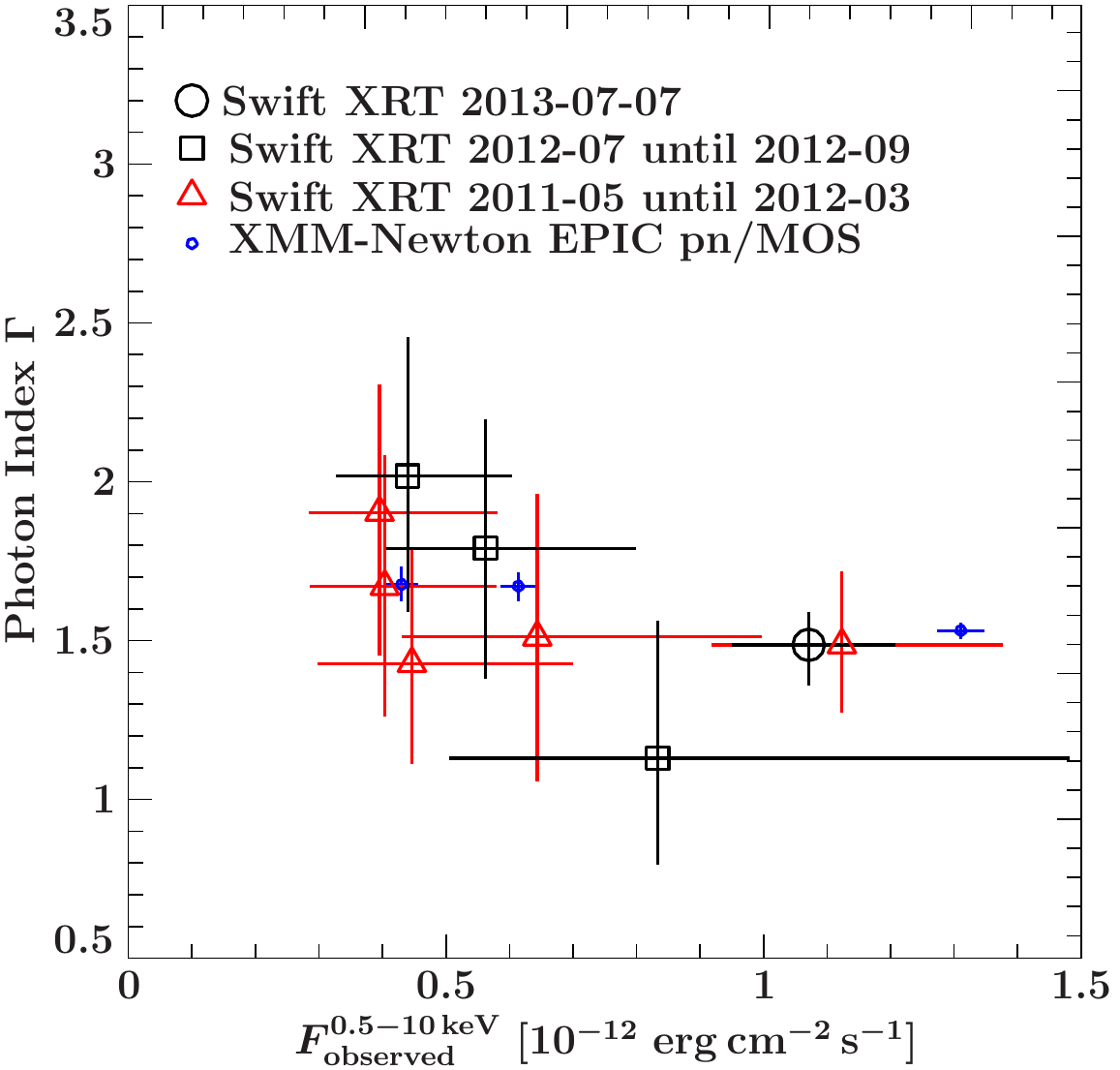} 
  \end{minipage}
  \caption{Hardness in the X-ray data of \pks{}.\textsl{Left panel:} Unabsorbed 2--10\,keV flux as a function of the 0.5--2\,keV flux. \textsl{Right panel:} Photon index of the absorbed powerlaw as a function of the unabsorbed 0.5--10\,keV flux.}
  \label{hdr}
\end{figure*}

To investigate whether variations in the source brightness are driven by the flux or changes in the spectral shape, Fig. \ref{hdr} shows the 2--10\,keV flux (hard flux, hereafter) plotted against the 0.5--2\,keV flux (soft flux, hereafter). A linear correlation suggests that the variations are driven by variations of the flux, instead of the spectral shape. The data are highly correlated (Spearman's rank coefficient $\rho=0.66$, p-value$=0.02$; Pearson's correlation coefficient $r=0.81$, p-value$=0.002$) but the large scatter and uncertainties of the \swift{} data do not allow us to draw any conclusions on the shape of the correlation and thus on the possibility of spectral variations. A linear fit to the well constrained data (\xmm{} data and \swift{} XRT 0003249200[5,6]) shows a negative off set of the hard flux, while the soft flux is zero. This indicates the existence of two emission components, one where flux variations are due to changes in the normalization of the initial powerlaw, and another soft component which in comparison appears to be non-variable. If the non-variable component has a different powerlaw index, this would imply a change of the spectral index when the non-variable soft component dominates the emission. The behavior of the photon index as a function of the total X-ray flux is also given in Figure \ref{hdr}. The \xmm{} observations in two low-state show a slightly steeper photon index ($\Gamma\sim 1.68\pm0.05$) compared to the 2004 \xmm{} and latest \swift{} observation ($\Gamma\sim 1.53\pm0.02$ and $\Gamma\sim 1.49\pm0.10$, respectively). These variations correspond to a change of 3.0-4.8 $\sigma$ and indicates a harder-when-brighter behaviour. More observations with high Signal-to-Noise ratio spectra, however, are needed to confirm this trend and the presence of two different emission components. 



\begin{thebibliography}{}
\bibitem{Abdo2009a}
Abdo, A.A. et al., ApJ \textbf{707}, L142 (2009)
\bibitem{BollerBrandtFabian1997MNRAS289}
Boller, Th. et al., MNRAS \textbf{289}, 393 (1997)
\bibitem{Cash1975}
Cash,W., ApJ \textbf{228}, 939 (1979)
\bibitem{Drinkwater}
Drinkwater, M.J., MNRAS \textbf{284}, 85, (1997)
\bibitem{Gallo2004}
Gallo, L.C., et al., MNRAS \textbf{370}, 245 (2006)
\bibitem{Kalberla}
Kalberla, P.M.W., et al., A\&A  \textbf{440}, 775 (2005)
\bibitem{Ohja2010}
Ojha, R. et al., A\&A \textbf{519}, A45 (2010)
\bibitem{Oshlack2001}
Oshlack, A.Y.K.N., Webster, R.L. \&  Whiting, M.T., ApJ \textbf{558}, 578 (2001)
\bibitem{Osterbrogge}
Osterbrogge, D.E., PNAS \textbf{75}, 540 (1978)
\end{thebibliography}
\end{document}